\documentclass[sigconf, preprint]{acmart}

\usepackage{array}

\usepackage{mdwmath}
\usepackage{mdwtab}

\usepackage{flushend}           
\usepackage{fixltx2e}

\usepackage{url}
\usepackage[utf8]{inputenc}
\usepackage{times}

\usepackage{versions}

\usepackage[capitalize]{cleveref}

\usepackage[caption=false,font=footnotesize]{subfig}

\usepackage{paralist}           

\usepackage{algorithmic}       
\usepackage{listings}           
\usepackage{moreverb}           

\usepackage{fancyvrb}           

\usepackage[gen]{eurosym}       
\usepackage{textcomp}           

\usepackage{pifont}             

\usepackage{units}              

\usepackage{booktabs}
\usepackage{ctable}             



\usepackage{rotating}           

\newcolumntype{L}[1]{>{\raggedright\let\newline\\\arraybackslash\hspace{0pt}}p{#1}}
\newcolumntype{C}[1]{>{\centering\let\newline\\\arraybackslash\hspace{0pt}}p{#1}}
\newcolumntype{R}[1]{>{\raggedleft\let\newline\\\arraybackslash\hspace{0pt}}p{#1}}

\newcommand{\PreserveBackslash}[1]{\let\temp=\\#1\let\\=\temp}

\usepackage[normalem]{ulem}     

\usepackage{fancyhdr}           
\usepackage{xcolor}
\usepackage[]{draftwatermark}   
\SetWatermarkText{PREPRINT}
\SetWatermarkColor[gray]{0.9}
\SetWatermarkAngle{45}
\SetWatermarkScale{1}
\newcommand{\preprintmsg}{This an authors' preprint.  Please cite as: John Noll, Mohammad Abdur Razzak, and Sarah Beecham. 2017. Motivation and Autonomy in Global Software Development: An Empirical Study. In Proceedings of the 21st International Conference on Evaluation and Assessment in Software Engineering (EASE'17). Association for Computing Machinery, New York, NY, USA, 394–399. DOI:https://doi.org/10.1145/3084226.3084277}
\newcommand{\preprinttitle}[1]{%
  \fancypagestyle{firstpagestyle}{%
    \fancyhf{}
    \fancyhead[L]{\vspace{-20pt}\footnotesize{\textcolor{gray}{#1}}}
  }
}
\newcommand{\preprintfoot}[1]{%
    \fancyfoot[L]{\vspace{10pt}\footnotesize{\textcolor{gray}{#1}}}
}

\usepackage{xspace}

\newcommand{\GSDlong}{Global Software Engineering\xspace}

\preprinttitle{\preprintmsg}
\preprintfoot{\preprintmsg}

\begin{document}
\copyrightyear{2017}
\acmYear{2017}
\setcopyright{acmcopyright}
\acmConference{EASE'17}{June 15-16, 2017}{Karlskrona, Sweden}
\acmPrice{15.00}
\acmDOI{http://dx.doi.org/10.1145/3084226.3084277}
\acmISBN{978-1-4503-4804-1/17/06}

\begin{abstract}

  Distributed development involving globally distributed teams in
  different countries and timezones adds additional complexity into an
  already complex undertaking.  This paper focuses on the effect of
  global software development on motivation. Specifically, we ask,
  what impact does misalignment between needed and actual autonomy
  have on global team motivation?

  We studied members of two distributed software development teams with different
  degrees of distribution, both following the Scrum approach to
  software development.  One team's members are distributed across
  Ireland, England and Wales; the other has members in locations
  across Europe and North America.  We observed the teams during their
  Scrum ``ceremonies,'' and interviewed each team member, during which
  asked we asked team members to rate their motivation on a 5 point ordinal scale.

  Considering both the reported motivation levels, and qualitative
  analysis of our observations and interviews, our results suggest
  that autonomy appears to be just one of three job aspects that
  affect motivation, the others being competence and relatedness.  We
  hypothesize that (1) autonomy is a necessary but not sufficient
  condition for motivation among experienced team members, and (2)
  autonomy is not a motivator unless accompanied by sufficient
  competence.
\end{abstract}

\keywords{
Global Software Development; 
Motivation; 
Empirical Software
Engineering; 
Autonomy; 
Self-Determination Theory; 
Career Anchors;
Agile Software Development; 
Scrum
}

\title[Motivation and Autonomy in Global Software Development]{Motivation and Autonomy in Global Software Development: An Empirical Study}

\author{John Noll}
\affiliation{%
  \institution{University of East London}
  \streetaddress{University Way, London, E16 2RD, UK}
}
\email{j.noll@uel.ac.uk}

\author{Mohammad Abdur Razzak}
\affiliation{%
  \institution{Lero, the Irish Software Research Centre,\\University
    of Limerick, Ireland}
}
\email{abdur.razzak@lero.ie}

\author{Sarah Beecham}
\affiliation{%
  \institution{Lero, the Irish Software Research Centre,\\University
    of Limerick, Ireland}
}
\email{sarah.beecham@lero.ie}

\maketitle

\section{Introduction}\label{sec:intro}

Geographic separation, lack of timezone overlap, cultural differences
and different first languages among the team  -- collectively referred
to as Global Distance -- make the already challenging task of software
development even more complex~\cite{Noll_2010_Global}. 

Autonomy has been identified as an important motivator for Software Engineers~\cite{Beecham_2008_Motivation}. 
A mismatch between an individual's need for autonomy, and the degree
of autonomy she or he actually has, may have an
impact on motivation levels~\cite{Porter_1998_Differential,Ferratt_2003_Instrument,Deci_1985_Cognitive}.
In this study, we ask,

\emph{Does increased autonomy result in higher motivation among distributed developers?}

As part of a larger software process improvement study~\cite{Noll_2016_Global}, we looked at the motivation of members of two distributed project teams, that made a transition from plan-driven to Agile development. 
We interviewed each team member, asking them to rate their motivation before and after the introduction of Scrum~\cite{Schwaber_2002_Agile}.  
Scrum emphasizes ``self-organizing teams'' that decide among themselves the best way to achieve their objectives~\cite{Sutherland_2016_Scrum}.
As such, we expected that motivation would be higher after the introduction of Scrum due to higher autonomy.
We found, however, that the difference in motivation levels before and after the introduction of Scrum was slight.
There was, however, an apparent difference in the motivation levels of
experienced team members, which were lower than less experienced
members after Scrum was introduced.
We speculate that this is due to the absence of other motivators as
put forward by Deci and Ryan's Self-Determination
Theory~\cite{Deci_2008_Self}, related to the need to be part of a
team, and the need to balance autonomy and competence.
We conclude that, while autonomy is an important motivator, it is not sufficient on its own.

This paper is organized as follows: in the next section we give a brief background to motivation theory in a global context, and reflect on changing software engineer characteristics and environmental factors, which motivate our research question. 
In \cref{sec:method} we describe the case study, including our data collection and analysis methods.  
In \cref{sec:results} we present our qualitative and quantitative results. 
\cref{sec:discussion} discusses how our results address our research question. 
We conclude the paper in \cref{sec:conclusions} with a summary of our findings, limitations, and plans for future work.

\section{Background}\label{sec:background}

We view motivation as a social process that defines how people join, remain part of, and perform adequately in, a human organization \cite{Huczynski_1991_Organizational}.
In this paper, we draw on motivation theory to help explain the significance of autonomy in a global organizational context. 
Due to its complex nature, motivation tends to be overlooked in project management since it is difficult to measure and control \cite{Beecham_2008_Motivation}, yet motivation is shown to have an impact on the quality of work produced \cite{Boehm_1981_Software}, productivity~\cite{Law_2005_Effects}, and on employee retention \cite{Hall_2008_Impact}. 

\subsection{Motivation and GSD}
There are numerous theories that try to explain the conscious or unconscious decisions people make to expend effort or energy on a particular activity \cite{Petri_2012_Motivation}. These theories provide insight into what motivates software engineers to engage fully in their tasks, commit to the organization's goals, and produce higher quality software \cite{Beecham_2008_Motivation}, and stimulate innovation \cite{Frey_2001_Successful}. Conversely, a demotivated workforce can lead to project failure \cite{Verner_2014_Factors}. Of particular relevance to this study is Self-Determination Theory, in which Ryan and Deci \cite{Ryan_2000_Self} postulate that to be self-motivated, employees require three basic psychological needs to be satisfied: 
\begin{inparaenum}
\item \emph{relatedness}, a feeling of connection to other members of a group;
\item \emph{competence}, having enough skills and expertise to carry out the job; and
\item \emph{autonomy}, the freedom to make decisions and do what's necessary, without undue constraints.
\end{inparaenum}

Some of the issues introduced by Global Software Development (GSD) \cite{Noll_2010_Global} may be addressed by meeting the motivational needs of software engineers. For example, GSD projects have been shown to suffer from high staff turnover \cite{Holmstrom_2006_Global,Ebert_2008_Managing}; 
conversely, high levels of motivation can have a positive effect on staff retention \cite{Hall_2008_Impact}.

Research has identified 
an inventory of 32 motivation factors for software
engineers \cite{Beecham_2008_Motivation,Francca_2011_Motivation,
  Fraias_2012_Motivational, Beecham_2015_What}. Among these are
problem solving, team working, change, challenge, benefit, and \emph{autonomy}.

\subsection{GSD impact on software engineer characteristics}

Of the many software engineer characteristics identified in the literature, ``growth oriented,'' ``introverted,'' and ``need for independence'' were the most cited, which indicates these occur across many different contexts. 
The view that software engineers are introverted reflects findings from the many studies coming from Couger and Zawacki and colleagues, that began in the 1980's, who measured the ``social needs strength'' of engineers \cite{Couger_1980_Motivating} in their Job Diagnostics Survey. This view is not universal, as some studies characterize software engineers as sociable people \cite{Beecham_2008_Motivation}.

Although some research suggests that the needs of a global software
engineer are similar to those of the general population of engineers
\cite{Khatib_2013_Role},  Beecham and Noll \cite{Beecham_2015_What}
speculate 
that the characteristics of global software engineers may be
changing; they observed that a software engineer working in a
distributed team is not de-motivated by the need to
travel or work anti-social hours, and is tolerant of work/life
imbalance.  They also found that engineers working in GSD
did not cite autonomy as important~\cite{Beecham_2015_What}.  
This distinguishing feature may be due to personality and ``individual
differences in their tendencies toward autonomous functioning across
specific domains and behaviors \cite{Ryan_2006_Self}.''  

\subsection{Motivation and Agile}

Traditionally, GSD has followed a plan-driven, structured, waterfall
approach, where tasks are allocated according to where they appear in
the software lifecycle \cite{Estler_2014_Agile}.  It was considered that
Agile methods, envisaged for \emph{small projects} and \emph{co-located
  teams} \cite{Kahkonen_2004_Agile, Abrahamsson_2009_Lots}, would be a
poor fit for GSD because Agile and distributed development
approaches differ significantly \cite{Ramesh_2006_Distributed}. Agile
methods tend to rely on informal processes to facilitate coordination,
whereas distributed software development relies on formal mechanisms.
Yet, there is a growing trend for companies engaged in GSD to adopt
Agile methods~\cite{Hanssen_2011_Signs, Fitzgerald_2013_Scaling}. Adopting Agile practices such
as short iterations, frequent builds, and continuous delivery all pose
challenges to configuration management and version management
\cite{Paasivaara_2006_CouldGSD}.  But, practices such as \emph{short
  iterations} increase transparency of work-in-progress, and
provide a big picture of project progress to stakeholders
\cite{Paasivaara_2004_UsingIterative}. However, setting up an Agile
team is usually motivated by benefits such as increased productivity,
innovation, and employee satisfaction
\cite{Smite_2010_FundamentalsAgile}.  

Introducing Agile methods can change the culture in a
company; developers need to have more autonomy as well as
decision-making power to implement  Agile practices in global
software environment \cite{Fowler_2006_Using}.  Through frequent communications
and meetings (i.e; daily stand-ups), Agile team members can motivate and
influence each other's behaviour \cite{Das_2001_Trust}; but little is
known about motivation in an Agile context
\cite{Beecham_2007_Does, McHugh_2010_Motivating}.

\subsection{Autonomy--still an important factor?}

We define autonomy as a feeling of independence, freedom and control (or self-determination) \cite{Deci_1985_Intrinsic}.
Autonomy has been identified in earlier studies as an important motivator for software engineers \cite{Beecham_2008_Motivation}, and is also a core concept in Self-Determination Theory \cite{Ryan_2000_Self,Gagne_2005_Self, Deci_1985_Intrinsic}. Ryan and Deci \cite{Ryan_2006_Self} reason that the more autonomy one feels, the more intrinsically motivated one becomes. 
It might be that the global software engineer profile is changing: there appears to be less interest in those factors that can act as barriers to motivation~\cite{Beecham_2015_What}. 
This may reflect Deci and Ryan's Cognitive Evaluation Theory (CET) \cite{Deci_1985_Cognitive} that specifically addresses social and environmental external factors that facilitate or undermine intrinsic motivation. Taking this argument forward, and given that many environmental factors are inevitable when working in GSD (such as having to meet colleagues virtually, fitting in with hours of remote colleagues in different timezones, and travel), Beecham and Noll suggest that those engineers that remain working in globally distributed teams for the long term, are resilient to the demotivating factors that are inherent in GSD.

In this study, we focus on one factor, and ask whether autonomy affects motivation of software engineers working in Global Software Development projects. Our research question is expressed as, \emph{``Does increased autonomy result in higher motivation among distributed developers?''}
We address this question through conducting a case study with a new and different set of engineers, also engaged in GSD; the approach is described in the next section.

\section{Method}\label{sec:method}
%

We conducted interviews with members of two distributed software development teams: the first (Team A) comprising six members located in the U.K. and Ireland, and the second (Team B) with nine members in Ireland and North America. 

The company we studied, which we shall call PracMed, is a medium-sized Irish-based software company that develops practice management software for health care professionals. PracMed employs approximately seventy staff members in its software development organization, including support and management staff. PracMed's annual sales approach \euro{} 20 million, from customers across Europe and North America.

The units of analysis of our study are the individuals that form two project teams.
Each project team focuses on  different aspects of PracMed's business, and exposes its members to different levels of Global Distance~\cite{Noll_2010_Global}. 
TeamA is responsible for maintaining the software that forms the core of PracMed's product line.
They also maintain and enhance the retail product for the Irish, British, Canadian, and Mexican markets. Finally, they perform maintenance on a legacy product resulting from an acquisition that also brought four of TeamA's team members to the company.
Two of TeamA's members work primarily from home in England; the other members are distributed equally between the head office in Ireland and an office in Wales.

TeamB's responsibility is to tailor the company's product for a large customer in North America. The nine members of TeamB are distributed over three countries in two continents, with up to eight hours difference in timezones between locations.

At the time we began our observations, both teams had been following
the Scrum~\cite{Schwaber_2002_Agile} software development method for
approximately six months.

\cref{tab:team-composition} ``Team Composition'' shows the distribution of members of both teams. 
In both teams, the Project Manager also plays role of Scrum Master.

\begin{table}
\caption{Case study team composition.}\label{tab:team-composition}
\centering
\begin{tabular}{lp{.23\columnwidth}clp{.23\columnwidth}c}
\toprule
\multicolumn{3}{c}{\textbf{TeamA}} & \multicolumn{3}{c}{\textbf{TeamB}}  \\                
\midrule
\textbf{Country} & {\textbf{Role}} & \textbf{\#} & \textbf{Country} & {\textbf{Role}} & \textbf{\#} \\ 
\midrule
Ireland          & Developer       & 2           & Ireland          & Product Owner   & 1           \\ 
Wales            & Scrum Master    & 1           &                  & Developer       & 3           \\ 
                 & Product Owner   & 1           &                  & QA              & 1           \\ 
England          & Sr. Developer   & 1           & Canada           & Scrum Master    & 1           \\ 
                 & QA              & 1           &                  & Product Owner   & 1           \\                
                 &                 &             &                  & Developer       & 1           \\                
                 &                 &             & USA              & Sr. Developer   & 1           \\ 
\midrule
                 & Total           & \textbf{6}  &                  & Total           & \textbf{9}  \\
\bottomrule
\end{tabular}
\end{table}

Two of the authors acted in a participant-observer role by sitting in on each team's Scrum ``ceremonies.'' 
One of us observed TeamA, daily, from November, 2015 to June, 2016;
another of us observed TeamB, daily, from January, 2016 to March,
2017. We sat in on daily standups, sprint planning, backlog grooming,
and sprint retrospectives. Due to the fact that the team members are
distributed across Europe and North America, the observations were
made by joining the video conference session for each ceremony. 

The observers also conducted semi-structured interviews of each member
of the team he was observing, according to an interview protocol
\cite{Beecham_2017_Lean}\footnote{Available at \url{http://www.lero.ie/sites/default/files/Lero_TR_2017_02_Beecham_Noll_Razzak-Lean\%20Global\%20Project\%20Interview\%20Protocol.pdf}}. 
During the interviews, all members of both teams were asked to describe
their backgrounds, roles on the team, and also the  development
processes from their point of view.  Each
interviewee was also asked to rate his or her motivation before and
after the introduction of Scrum, on a five point ordinal scale
comprising
\begin{inparaenum}
\item ``definitely low,'' \item ``somewhat low,'' \item ``neither low
  nor high,'' \item ``somewhat high,'' and \item ``definitely high''
\end{inparaenum}.  Since the interviews were semi-structured, each
participant was encouraged to elaborate on their answers, and in
particular to explain why they rated their motivation as they did.

After transcribing the interviews, the motivation ratings were
tabulated to compare team motivation before and after the introduction
of Scrum.  Subsequently, we examined the interviewees' explanations
for their ratings in order to understand the reasons behind the
values (as shown in \cref{tab:motivation-before-after-scrum}).  The results are described in the next section.

\section{Results}\label{sec:findings}\label{sec:results}

We first present results of team members' self-reported motivation. 
As mentioned in the previous section, at the end of each interview, interviewees were asked to rate their motivation on a five-point ordinal scale; 
\cref{tab:motivation-before-after-scrum} summarizes the results
(note: before the introduction of Scrum, there was a total of twelve team members; three additional members were hired after Scrum was introduced).

The reported motivation levels range from ``neither low nor high'' to
``very high'' motivation, both before and after the introduction of Scrum; no team member reported low motivation.  
The most common motivation level before the introduction of Scrum was ``Neither low nor high,'' while after the introduction of Scrum, motivation levels were evenly distributed among ``Neither low nor high,'' ``Somewhat high,'' and ``Definitely high,'' with some team members reporting higher motivation, others reporting lower, and some reporting no difference.
This suggests that introducing Scrum had a positive effect on team member motivation (see~\cref{tab:motivation-before-after-scrum}).

\begin{table}
\centering
\caption{Individual motivation before and after the introduction of Scrum.}
\label{tab:motivation-before-after-scrum}
\begin{tabular}{lcccccc}
\toprule
& \multicolumn{3}{c}{\textbf{Rating}} &&\\
\textbf{Scrum stage} & \textbf{3} & \textbf{4} & \textbf{5} & \textbf{Total} & \textbf{Median} & \textbf{Mode} \\
\midrule
before               & 5          & 4          & 3          & 12             & 4               & 3             \\ 
\midrule
after                & 5          & 5          & 5          & 15             & 4               & 3, 4, 5       \\ 
\bottomrule
\end{tabular}
\end{table}

Experience does seem to affect motivation: the highest motivation scores were reported by the team members with less than ten years experience (\cref{tab:team-member-motivation-by-experience}).
Comparing these less experienced developers to their peers with ten or more years experience,
there appears to be slightly higher motivation among experienced developers before the introduction of Scrum.
However, after the introduction of Scrum, the \emph{less experienced} developers had somewhat higher motivation (\cref{tab:team-member-motivation-by-experience}).

\begin{table}
\caption{Team member motivation by experience.}
\label{tab:team-member-motivation-by-experience}
\centering
\begin{tabular}{p{.13\columnwidth}p{.13\columnwidth}cccccc}
\toprule
\textbf{Experi-} & \textbf{Scrum} & \multicolumn{3}{c}{\textbf{Rating}} &  &  &  \\
\textbf{ence} & \textbf{Stage} & \textbf{3} & \textbf{4} & \textbf{5} & \textbf{Total} & \textbf{Median} & \textbf{Mode} \\
\midrule
$<$10 & before & 2 & 1 & 1 & 4 & 3.5 & 3      \\
years & after  & 0 & 4 & 3 & 7 & 4   & 4      \\
\midrule
10+   & before & 3 & 3 & 2 & 8 & 4   & 3 \& 4 \\ 
years & after  & 5 & 1 & 2 & 8 & 3   & 3      \\
\bottomrule
\end{tabular}
\end{table}

\section{Discussion}\label{sec:discussion}

In Agile software development, individuals and teams are given greater autonomy and
decision-making power in monitoring and selecting tasks, as compared to
traditional, plan-driven development \cite{Fowler_2006_Using}.  In
particular, one of the hallmarks of Scrum is the concept of
``self-organizing teams,'' which means that Scrum teams are allowed
(and even required) to figure out for themselves how to achieve their
objectives~\cite{Sutherland_2016_Scrum}.  As such, Scrum teams
exercise a high degree of autonomy.  So, why didn't we see improved
motivation after the introduction of Scrum?

For this neutral result we need to focus on the more experienced engineers, the majority of whom registered a neutral ``neither low nor high'' level of motivation after the introduction of Scrum. Given that high motivation among individuals is desirable, we view the fact that one third of the sample (5 members) were not highly motivated as sub-optimal. 

One possible reason for the lower motivation among the more experienced developers might be due to these developers having less autonomy than would normally be found in an Agile environment.
Our findings showed hints of issues concerning autonomy in responses
from one of the  more experienced developers:

\emph{In my own point of view if [Product Manager] says, this ticket takes five days, then and as a developer if I think it will take ten days then  probably I shouldn't say ten days because other people may think I am taking more time. The reality is everyone takes a longer time than the initial estimation.}

In contrast, the more junior members
were comfortable with their dependence on senior
developer inputs in the planning.
One of TeamB's junior offshore developers reported that all of his code had to be
reviewed by a senior developer in Dublin, suggesting a lack of autonomy.
Yet the same developer said that ``checks and balances'' and
``more communication,'' which are hallmarks of Scrum, result in a
``better product at the end of the sprint,'' suggesting that he did not
perceive a mismatch between his desired and actual autonomy.
Another interviewee noted that the lack of input in planning
affected sprints, which he suggested could be addressed through
more support from the senior developer, rather than working alone. 

This was echoed by a third developer:

\emph{If you look at our velocity chart you can see we haven't met
  most of the commitments. So, I think [Senior Developer] should estimate.}

This reflects a healthy attitude as noted in earlier work with a high
performing Agile team where decisions were made by consensus, and when
asked about what drives \emph{down} performance, the high performing
team members responded ``developers wanting to do things the way they
want to and not listen to anyone else \cite{Beecham_2007_Does}.'' 
Developers in our sample were exhibiting similar behaviour to those
developers in the high performance team. 

But this view is not universal: 

\emph{\ldots I need to get approval from too many people--one person says
`okay good, check with X' or `no [it's] rubbish take it back.'  It is too
difficult to satisfy that many people.  That means I would start
something with [Product Manager] and he says `that's good.'  Then
[Technical Lead] says `tweak over here and I am happy with it.'  Then I
go to the CTO and he said, `no take it off.'  That is big problem for
me initially at least. The amount of rework I had to do on everything
was tremendous.}

Nevertheless, this new team member rated his motivation as ``definitely high.''

Senior team members seem to have a different perspective.
One senior person with over 15 years domain experience rated motivation as ``neither low nor high,''
attributing this to lack of experience and training in his new role as  Product Owner:

\emph{In my previous job my motivation was quite high. But, in this particular project it's neither low nor high because of the learning curve. And, it's been very difficult to wrap my head around the roles itself. For that reason, it's been quite demoralizing actually when you don't fully understand your job properly. You don't really know 100\% what and why you're supposed to do. It could be quite demoralizing and it does affect your motivation.}

In contrast, a senior developer in the home office rated his motivation as ``definitely high,'' citing intellectual challenge as the reason:

\emph{To be honest as long there is new stuff to do \ldots [the process] doesn't matter. So, it's very high.}

It appears we are seeing examples of Ryan and Deci's Self-Determination Theory~\cite{Ryan_2000_Self} in action (see
\cref{sec:background}).  The motivated individuals, whether junior or experienced, appear to have
sufficient skills (competence),
support from other team members (relatedness), and autonomy, to carry out their
tasks.  

Unmotivated individuals appear to be lacking one of these
aspects: the Product Owner with 15 years domain experience but no
experience as a Product Owner is lacking competence in this specific
role, despite having autonomy.  

According to Ferratt et al \cite{Ferratt_2003_Instrument}, Shein's
\cite{Schein_1996_Career} eight \emph{career anchors} can explain how
motivation needs to be finely tuned to each individual.  A career
anchor is a person's self-perceived talents and abilities, basic
values, motives and needs as related to his or her career.  For
example ``people whose anchors are security/stability should prefer
longer-term embedded, more secure arrangements. Those who value
autonomy/independence might value arrangements with greater
discretion \cite{Schein_1996_Career}.'' Also, Sumner and colleagues found
four career anchors or orientations were prevalent among IT
professionals: creativity, autonomy, identity, and
variety~\cite{Sumner_2005_Career}.  So autonomy alone may not be
sufficient for a person to be motivated, and might not even be motivating
at all.

Further, Prasad et al found that employment
arrangement characteristics and IT employment arrangement satisfaction
vary according to individual work value profiles
\cite{Prasad_2007_One}.
As such, it may be that an individual's need for
autonomy forms part of a work values profile which can be viewed as an
individual psychological construct.  Porter anticipated this result by
suggesting managers ``offer increasing responsibility and autonomy as
employees feel they are ready. \cite{Porter_1998_Differential}''

\begin{figure}
  \centering
  \includegraphics[width=.6\columnwidth]{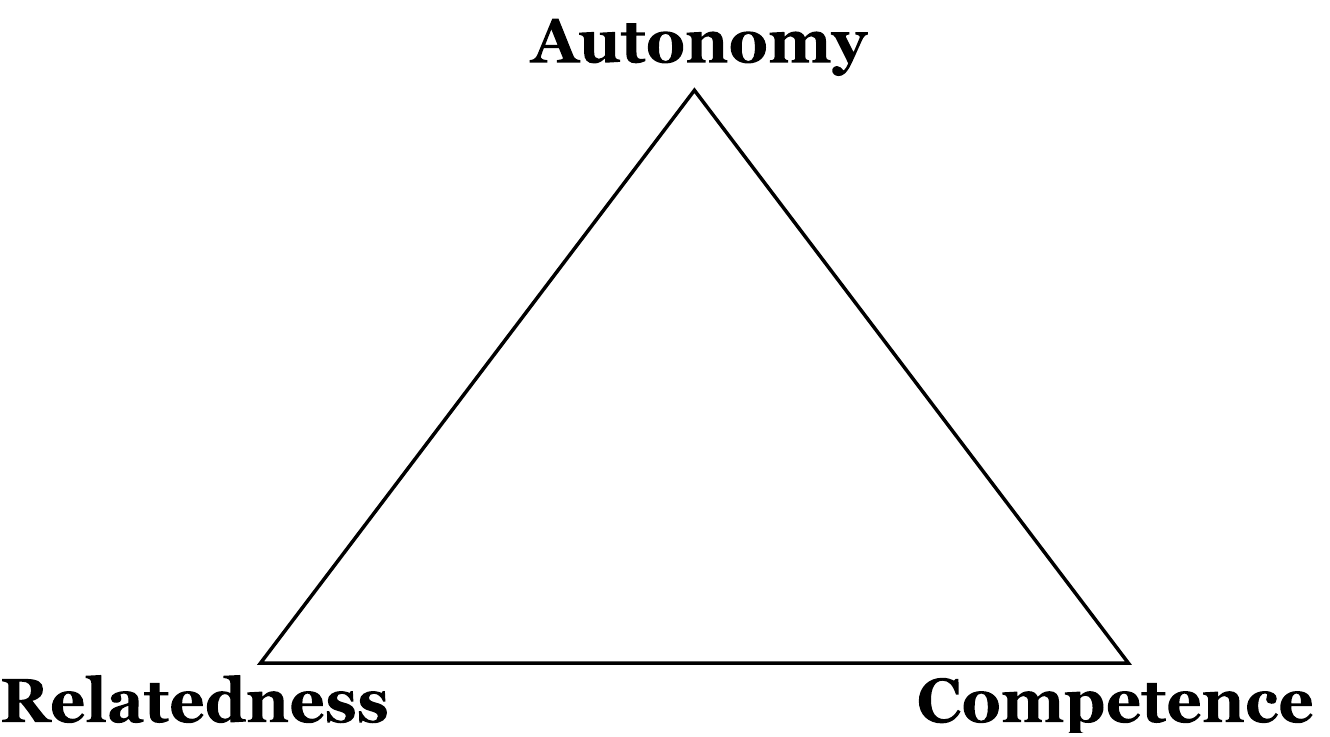}
  \caption{Self-determination Theory Psychological Constructs.}
  \label{fig:self-determination-triangle}
\end{figure}

Returning to our research question, \emph{``Does increased autonomy result in higher motivation among distributed developers?''},
it seems that autonomy  needs
to be matched by competence and relatedness (\cref{fig:self-determination-triangle}).  Perhaps the more
experienced software engineers in our sample lacked autonomy
commensurate with their competency (skills and experience).
Similarly, team members new to their roles may have had \emph{too
  much} autonomy, as they lacked the skills and experience to make
decisions without consulting others.  Finally, motivated junior
developers enjoyed the support they received from senior colleagues,
satisfying (in part) their need for relatedness; the associated lack
of autonomy was not seen as de-motivating because these junior
developers had not yet achieved sufficient competency in the specific
context to act autonomously.

\subsection{Limitations}
This study has a number of limitations that might threaten the validity of the conclusions.

Our five-point measure of motivation is quite coarse, and was applied only once during the study period. 
As a construct it seems reasonable, but does not delve into the reasons why a team member might choose a particular value.
Also, we might find interviewees report different levels of motivation at different times, being influenced by other factors. 

Also, we did not have a direct measure of autonomy; rather, we assumed that Scrum would give teams and team members higher autonomy than a plan-driven approach.  
It is possible that PracMed's particular approach to implementing Scrum might result in less autonomy than would normally be the case, although we have no evidence to suggest this is the case.

The individuals we studied were members of teams that were transitioning from plan-driven to Agile development: 
at the time we began our observations, each team had been following the Scrum approach for approximately six months, which means their Scrum adoption might have been incomplete, and they may not have had enough time to absorb or appreciate the new ways of working.  
However, our observations indicated the contrary: both teams seemed comfortable with the various Scrum ceremonies, and, aside from issues stemming from \GSDlong, both teams appeared to be working effectively as Scrum teams.

Our sample is small as we focused on just two teams of a single company, comprising a total of fifteen members. 
As such, the sample is not large enough to detect statistically significant differences in motivation levels before and after introducing scrum. 
In any case, given that a single company is involved, we have not attempted to generalize our findings.  

We can, however, use our findings to generate hypotheses to motivate future research.  
Therefore, we propose the following hypothesis based on our observations:

\emph{\textbf{Hypothesis 1:} Autonomy can be de-motivating unless
accompanied by sufficient competence.}
As noted before (\cref{sec:background}), Ryan and Deci assert that autonomy and competence are
two of three psychological pre-requisites for
motivation~\cite{Ryan_2000_Self}.  As such, autonomy has to be matched
to competence to exercise that autonomy
effectively~\cite{Porter_1998_Differential}
This hypothesis states that the converse -- that autonomy without
competence can be de-motivating -- follows from our observation that
some team members who had substantial autonomy but lacked experience
in their specific roles were not highly motivated.

\emph{\textbf{Hypothesis 2:} Individuals who recognize they lack
competence in their current role will be motivated by relatedness
rather than autonomy.}
This hypothesis is a corollary to Hypothesis 1, and follows from
comments from inexperienced team members, who reported support from
senior members as positive aspects of their roles.

\emph{\textbf{Hypothesis 3:} Competent individuals will find a lack of autonomy de-motivating.}
This hypothesis follows from the observation that all of the team members who reported neither low nor high motivation after the introduction of Scrum had more than ten years of experience.
Also, our previous investigation into developer motivation in GSD indicated that senior developers do not consider autonomy to be as important as other factors, 
such as intellectual challenge, making a meaningful contribution, and creativity~\cite{Beecham_2015_What}.
We note that in both studies, the  personnel presumably had significant autonomy, by virtue of their being senior level staff (project managers, senior managers, and directors in the previous study, and members of Scrum teams in the current study); 
as such, it seems there are other factors that contribute to motivation, or lack thereof.

\section{Conclusions}\label{sec:conclusions}
%

Prior research has shown that software engineers who are motivated deliver
higher quality software \cite{Beecham_2008_Motivation},
are more innovative \cite{Frey_2001_Successful}, 
more successful \cite{Verner_2014_Factors}
and less prone to attrition 
\cite{Hall_2008_Impact}. Companies, including those with GSD teams,
are adopting Agile methods~\cite{Hanssen_2011_Signs}
in an effort to realize benefits such as increased
productivity, innovation, and employee satisfaction 
\cite{Smite_2010_FundamentalsAgile}. However, Agile methods were
originally designed for small, co-located teams 
\cite{Abrahamsson_2009_Lots,Kahkonen_2004_Agile}, and require significant
autonomy to be fully deployed \cite{Fowler_2006_Using}.

Autonomy is also an important factor in motivation for software engineers 
\cite{Beecham_2008_Motivation},
and is linked to their intrinsic motivation \cite{Ryan_2006_Self}.
Self-Determination theory holds that autonomy is one of three psychological needs
which must be met \cite{Ryan_2000_Self}.

Because autonomy is a crucial component of Agile development, as well as an
important for software engineers' motivation, but potentially difficult
to satisfy within the context of GSD, we examined the extent to which autonomy
affects the motivation of members of two different GSD teams within the same
organization. Previous work on software engineers in GSD has shown that they
may become resilient to typical demotivating factors which are an unavoidable
component of GSD \cite{Beecham_2015_What}, 
leading us to ask if increased autonomy significantly affects global 
software developer motivation in an Agile environment.

We observed the two teams over a period of months, and interviewed their members. 
We asked each team member to rate the level of motivation that he or she felt, on a five point ordinal scale, both before and after the introduction of Agile methods to their development process.

So far, we have found little evidence to suggest there is a difference
in motivation between members of teams that are distributed but
located in the same general vicinity, and those in teams who are
distributed across continents. 
What differences exist seem to be more related to experience, a
sense of being part of a team, and factors such as intellectual
challenge and contributing to a valuable product, that were
previously identified as contributing to motivation in global software development \cite{Beecham_2015_What}.

In future work, we plan to administer the motivation survey used by Beecham and Noll to the members of TeamA and TeamB to try to understand the reasons behind their particular motivation ratings.

\section{Acknowledgments}\label{sec:acknowledgments}
We thank the members of TeamA and TeamB for their generous and thoughtful collaboration on this study, and PracMed, for allowing us to study their software development efforts. This work was supported, in part, by Science Foundation Ireland grants 10/CE/I1855 and 13/RC/2094  to Lero - the Irish Software Research Centre (\url{www.lero.ie}).

\bibliographystyle{ACM-Reference-Format}
\bibliography{top} 

\appendix


\end{document}